\documentclass[aps,prd,a4paper,preprintnumbers,amsmath,amssymb,superscriptaddress,twocolumn,floatfix,showpacs,showkeys,nofootinbib]{revtex4}

\newcommand{\be}{\begin{equation}} \newcommand{\ee}{\end{equation}}
\newcommand{\ba}{\begin{eqnarray}} \newcommand{\ea}{\end{eqnarray}}
\newcommand{\bc}{}

\usepackage{amsfonts} \usepackage[dvips]{graphicx}
\DeclareGraphicsExtensions{.ps}

\begin{document}

\title{The cosmological behavior of
Bekenstein's modified theory of gravity}
\author{F.~Bourliot} \affiliation{Astrophysics, University of Oxford,
Denys Wilkinson Building, Keble Road, Oxford OX1 3RH, UK}
\author{P.G. Ferreira} \affiliation{Astrophysics, University of
Oxford, Denys Wilkinson Building, Keble Road, Oxford OX1 3RH, UK}
\author{D. F. Mota} 
\affiliation{ Institute for Theoretical Physics, University of Heidelberg, 69120 Heidelberg,Germany }
\affiliation{Institute of Theoretical
Astrophysics, University of Oslo, Box 1029, 0315 Oslo, Norway}
\affiliation{Perimeter Institute, Waterloo, Ontario N2L 2Y5, Canada}
\author{C. Skordis} \affiliation{Perimeter Institute, Waterloo,
Ontario N2L 2Y5, Canada}

\date{\today}

\begin{abstract}
We study the background cosmology governed by the Tensor-Vector-Scalar
theory of gravity proposed by Bekenstein. We consider a broad family
of functions that lead to modified gravity and calculate the
evolution of the field variables both numerically and analytically. We
find a range of possible behaviors, from scaling to the late time
domination of either the additional gravitational degrees of freedom or the
background fluid.
\end{abstract}

\keywords{Cosmology: Theory} \pacs{98.80.-k,98.80.Jk}

\maketitle

\noindent

\section{Introduction}
Bekenstein has proposed a covariant and relativistic theory of gravity~\cite{Bek} which 
is to meant to provide for an alternative to dark matter. The theory is 
called TeVeS (which stands for Tensor-Vector-Scalar)
and can have the Bekenstein-Milgrom Aquadratic Lagrangian~\cite{BM} nonrelativistic limit,
which provides for  Milgrom's  Modified Newtonian Dynamics~\cite{Milgrom}
(MOND) as an explanation of galactic rotation curves.

 MOND has had a number of successes at fitting data on galactic scales, less so
on the scale of clusters. As shown by Sanders \cite{Sanders}, a
universe governed by MOND still requires a small fraction of massive
neutrinos (corresponding to a neutrinos mass of approximately $2$ eV)
to account for the dynamics of clusters. More recently, the data
obtained in Clowe {\it et al} \cite{Clowes} indicates that some
extra mass would be needed in the ``Bullet'' cluster to account for
the mismatch between X-ray and weak lensing reconstructions of its
mass profile. Again, this can be shown, in the context of MOND, to be
resolved with a small fraction of massive neutrinos, as claimed by
Sanders \cite{Zhao}. However none of these results need to be valid in TeVeS: 
the vector field which was shown to be extremely important for structure formation~\cite{DL},
is also expected to be important on the scale of clusters.

MOND is the nonrelativistic limit, in the regime of small acceleration, of 
TeVeS. This is exactly akin to Newtonian gravity being the nonrelativistic
limit of General Relativity. And in the same way, it is incorrect to
infer all properties about the relativistic theory from the nonrelativistic
regime. The most notable case is in gravitational lensing: the Newtonian
limit supplies the wrong answer if one wishes to study general relativity.
Hence results obtained in the MONDian regime do not necessarily extend
to TeVeS and in many situations one must explore the fully relativistic theory. One regime
where this is imperative is the large scale dynamics of the Universe.
 In this paper we try to infer general properties of a universe governed by TeVeS.

 TeVeS incorporates a dynamical scalar field $\phi$, two metric tensors and a unit-timelike vector field $A^{\alpha}$. 
 In the case of a homogeneous and isotropic space-time, 
the TeVeS field equations reduce to a
form similar to Friedman-Robertson-Walker (FRW) form. There have been 
several studies\cite{Bek,DSR,Hao,SMFB,DL} of the resulting background cosmology
with different choices of the TeVeS free function.
In this paper we extend the preliminary analysis of TeVeS cosmology to more general
settings.

The paper is organized as follows. In Sec. 2 we give the general TeVeS
equations, their applications to cosmology and the notation we will
use throughout the paper. In Sec. 3 we will present our numerical
results and we will discuss their behavior with some analytical
insights. We conclude in Sec. 4.

\section{Fundamentals of TeVeS cosmology}

\subsection{TeVeS action}
Let us first describe the basics of TeVeS theory. We use the
conventions of \cite{skordis}.

In TeVeS we have two tensor fields which are the two metrics $g_{\mu
\nu}$ and $\tilde{g}_{\mu \nu}$ mentioned above, a scalar field $\phi$
and a vector field $A^{\alpha}$. They can be linked through
\begin{equation} \label{chgtmetr}
g_{\mu \nu}=e^{-2\phi}\left(\tilde{g}_{\mu \nu}+A_{\mu}A_{\nu} \right)
-e^{2\phi}A_{\mu}A_{\nu}
\end{equation}

The total action of the system is
$S_{\text{sys}}=S_{g}+S_{s}+S_{v}+S_{m}$ where $S_{g}$ is the
traditional Einstein-Hilbert action built with the
\textit{Einstein frame} metric $\tilde{g}_{\mu\nu}$, $S_{s}$ is the action for $\phi$ and
$\mu$, $S_{v}$ is the action for the vector field and $S_{m}$ denotes
the action of ordinary matter. Explicitly we have
\begin{eqnarray*}
S_{g} &=& \frac{1}{16 \pi G} \int d^4x \sqrt{-\tilde{g}} \tilde{R} \\
S_s &=& -\frac{1}{16\pi G} \int d^4x \sqrt{-\tilde{g}} \left[\mu
\left( \tilde{g}^{\mu \nu}-A^{\mu}A^{\nu} \right)\phi_{,\alpha}\phi_{, \beta}+ V(\mu) \right]
\label{dforme}\\ S_v &=& -\frac{1}{32\pi G}\int d^4x
\sqrt{-\tilde{g}}\left[K_B
F^{\alpha\beta}F_{\alpha\beta}-2\lambda(A^{\mu}A_{\mu}+1)\right]\\
S_{m} &=& \int d^4x \sqrt{-g} L_m\left[g_{\mu \nu}, \chi^A,\nabla \chi^A \right]
\end{eqnarray*}
where  $G$ is the bare gravitational constant, $K_B$ is a dimensionless constant, $\lambda$ is  a 
Lagrange multiplier imposing a unit-timelike constraint on the 
vector field, $F_{\alpha\beta}=2 \nabla_{[\alpha}A_{\beta]}$ and $\mu$ 
is a nondynamical field with  a free function $V(\mu)$. 
The matter Lagrangian $L_m$ depends only on the \textit{matter frame} metric $g_{\mu\nu}$
and a generic collection of matter fields $\chi^A$.

Although the theory was designed to provide for a relativistic theory
of MOND, taken at face value this need not be the case. It all lies
within the choice of the function $V(\mu)$. The requirement that both
exact MOND and exact Newtonian limits exist within the theory puts
some constraint on the possible forms of $V(\mu)$. In particular these
two requirements fix the derivative of $V(\mu)$ to be of the form
\begin{equation}
 V'(\mu ) \equiv \frac{dV}{d\mu}(\mu) = -\frac{1}{16\pi\ell^2_B} \frac{\mu^2}{(1 -
 \mu/\mu_0)^m} f(\mu)
\end{equation}
where $\ell_B$ is a scale, $\mu_0$ a dimensionless constant and $f(\mu)$ is still an
arbitrary function, not related to the MOND or Newtonian limits, whose
only constraint is that $f(0) = f_0$ is nonvanishing. The $\mu^2$ in the
numerator is what is precisely required to have an exact MOND limit
(which is reached as $\mu\rightarrow 0$), whereas the factor $1 -
\mu/\mu_0$ is what is required to have an exact Newtonian limit (which
is reached as $\mu\rightarrow \mu_0$).  Bekenstein chooses the
simplest case, namely $m=1$, which we shall also do in this work.
The measured Newton's constant $G_N$ and MOND acceleration $a_0$ are
given irrespectively of the form of $f(\mu)$ or the power $m$, as
\begin{equation}
 G  = \frac{\mu_0 (1 - K_B/2)}{\mu_0 + 2 - K_B} \;  G_N
\end{equation}
and
\begin{equation}
  a_0 =  \frac{e^{\phi_c}}{\ell_B \sqrt{\pi}}\left( \frac{1}{2-K_B} + \frac{1}{\mu_0}\right) 
\end{equation}
where $\phi_c$ is the cosmological value of $\phi$ when the quasistatic system in question broke away from the cosmological expansion.

What remains is to choose the (still free) function $f(\mu)$. It turns
out that for cosmological models we must have $V'\ge0$. Therefore in
the $m=1$ case that we are considering, this cannot happen in the same
branch as for a quasistatic system, i.e. we need $\mu>\mu_0$ for
cosmology. (There is a very clever alternative considered in
\cite{ZF} with its cosmology studied in \cite{ZFcosmo}. We do not follow that approach here). In the case that
there is an single extremum in $V'$ for $\mu>\mu_0$ one has a further choice
imposed by the requirement that $V'$ be single valued in the branch
considered  : either use the branch from $\mu =
\mu_0$ up to the position of the extremum or the branch from the
extremum to infinity. Bekenstein makes the second choice, which is
what we will also do in this work. 

With the above considerations in mind, Bekenstein makes the choice
$f(\mu) = (\mu/\mu_0 - 2)^2$. We will therefore generalize this choice
as
\begin{equation}
\label{oury}
 f(\mu) = \sum_n c_n (\hat{\mu} - \mu_a)^n
\end{equation}
where $\hat{\mu} = \mu / \mu_0$, $\mu_a$ is a constant and $c_n$ a set of additional constants. We also let $n$ run over
negative values. The Bekenstein model is recovered as $\mu_a = 2$, $c_2 = 1$ and $c_{n\ne2} =0$.

Different functions have also been proposed especially by 
\cite{ZF} and  \cite{FamBin} and slight modifications to Bekenstein's toy model have
been given by \cite{Hao}.

\subsection{Homogeneous and isotropic cosmology}
For the purpose of this paper we will restrict ourselves to
homogeneous and isotropic spacetimes (see \cite{SMFB,skordis} for a
detailed investigation of the inhomogeneous case). We adopt the
Friedmann-Lema\^{i}tre-Robertson-Walker metric with the scale factor
$a$ for the matter frame metric and $b=ae^{\phi}$ for the Einstein
frame metric.  We choose a coordinate system such that the components
of the vector field take the form $A^{\alpha}=(1,0,0,0)$. The modified
Friedmann-equation becomes
\begin{equation} \label{fried}
3\tilde{H}^2=8\pi G e^{-2\phi} \left(\rho_\phi + \rho_X
\right)
\end{equation}
where $\tilde{H} = \dot{b}/b$ is the Einstein-frame Hubble parameter
with an overdot means a derivative with respect to the coordinate
time chosen and where we have defined a field density $\rho_\phi$ as
\begin{equation}
 \rho_\phi = \frac{1}{16\pi G} e^{2\phi} \left(\mu V'+V\right)
\end{equation}
The ordinary fluid energy density $\rho_X$ evolves as usual as
\begin{equation} \label{Bianchi}
\dot{\rho}_X=-3\frac{\dot{a}}{a} (1 + w) \rho_X
\end{equation}
where $w$ is the equation of state parameter of the fluid.
Relative densities $\Omega_i$ can be defined as usual as
\begin{equation}
  \Omega_i = \frac{\rho_i}{\rho_i + \rho_\phi}
\end{equation}

Variation of the action with respect to the field $\mu$ gives back the
constraint
\begin{equation}
\label{constraint}
\dot{\phi}^{2}=\frac{1}{2} V'
\end{equation}
which fixes $\mu$ as a function of $\dot{\phi}$.  

Finally  the scalar field evolves according to a system of two first
order equations given by 
\begin{equation}
    \dot{\phi}  =  -\frac{1}{2\bar{\mu}} \Gamma
\end{equation}
and 
\begin{equation}
\label{evol_gamma}
 \dot{\Gamma} + 3\tilde{H}\Gamma  = 8\pi G e^{-2\phi} \rho_X (1 + 3w).
\end{equation}

\section{Numerical Results}
\subsection{Bekenstein's toy model}

Let us first revise the results presented in \cite{SMFB}, i.e.  that
$\phi$ acts as a tracker field in the radiation, matter and $\Lambda$
eras.

In this  special case of the Bekenstein toy model,
the function $V'$ can be explicitly inverted analytically by setting
$q = 64\pi l_{B}^{2}\dot{\phi}^{2}/(3\mu_0)$, $r = q\sqrt{q^2/12 + 64/81}$
$s = (27q^{2}+128)/54$, $p_1 = (r + s)^{1/3}$,
$p_2 = \sqrt{9p_1+12+16/p_1} $, 
$ p_{3} = \left(9p_1^2-24p_1+16 \right)p_2 -54 q p_1$
 and one obtains an explicit expression for the nondynamical field $\mu$:
\begin{equation}
 \mu=\frac{1}{6} \sqrt{-\frac{p_3}{p_1p_2}} +\frac{1}{6} p_2+1
\end{equation}

\begin{figure}[htbp]
\begin{center}
\includegraphics[scale=0.4]{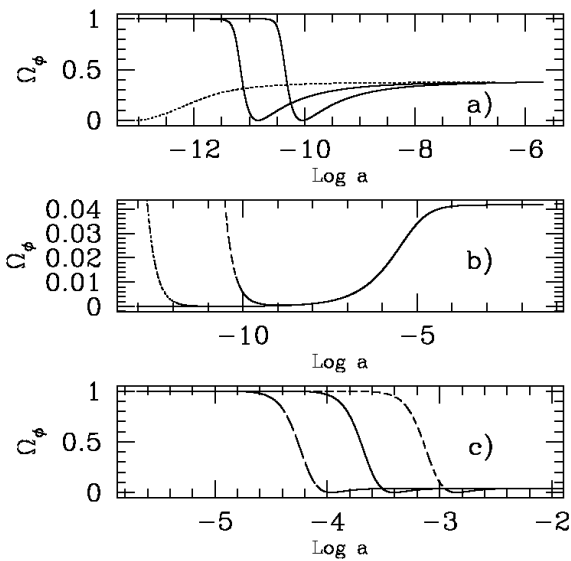}
\end{center}
\caption{ $\Omega_{\phi}$ as a function of $\text{log}(a)$. The units
for the scale factor are arbitrary. $\phi$ has an attractor which
correspond to tracking the three constituents independently of the
initial conditions. We illustrate this here, taking $\mu_{0}=4$ and
plotting three different histories in: $a)$ radiation era, $b)$ matter
era and $c)$ the $\Lambda$ era.  }
\label{fig1}
\end{figure}

In fig.\ref{fig1} we plot $\Omega_{\phi}$ as a function of
$\text{log}(a)$ for the different epochs in the Universe history:
panel $a)$ radiation era, panel $b)$ matter era and panel $c)$ the
$\Lambda$ era. As can be seen in the figure, for several different
initial conditions, $\phi$ has an attractor in the different epochs,
which leads to the tracking behavior.

\begin{figure}[htbp]
\begin{center}
\includegraphics[scale=0.4]{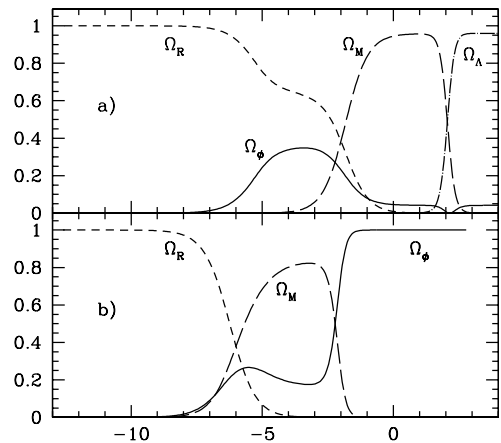}
\end{center}
\caption{Panel $a)$ represents Bekenstein's toy model, panel  and $b)$ is a
model with $c_{2},\ c_{-1}$ and $c_{-2}$ as non vanishing.  The
long dashed line represents $\Omega_{R}(t)$, the dot-dashed one is
$\Omega_{\Lambda}(t)$, the solid line is $\Omega_{\phi}(t)$ and the
short dashed line is for $\Omega_{M}(t)$.}
\label{fig2}
\end{figure}
In panel (a) of  fig.\ref{fig2} we plot the fractional energy densities $\Omega_{i}$
as a function of $\text{log}(a)$.  As can be seen from the figure,
$\phi$ synchronizes its energy density with the dominant component of
the Universe. As mentioned in \cite{SMFB}, during the radiation era
$\Omega_{\phi}$ reaches $\frac{3}{2 \mu_{0}}$ and during the matter
and $\Lambda$ eras the limit is $\Omega_{\phi}=\frac{1}{6 \mu_{0}}$.
>From panel (a) of fig.\ref{fig2} it is clear that the overall behavior of $\phi$ is
simply a sequence of the three tracking epochs.  This behavior is
similar to other general cosmological theories involving tracker
fields \cite{quintessence}.  

\subsection{Generalized function}

The behavior of $\phi$ depends on the form of the function $f(\mu)$ chosen. In
this section we investigate the main features which result from a
function given by equation (\ref{oury}).
\begin{figure}[htbp]
\begin{center}
\includegraphics[scale=0.4]{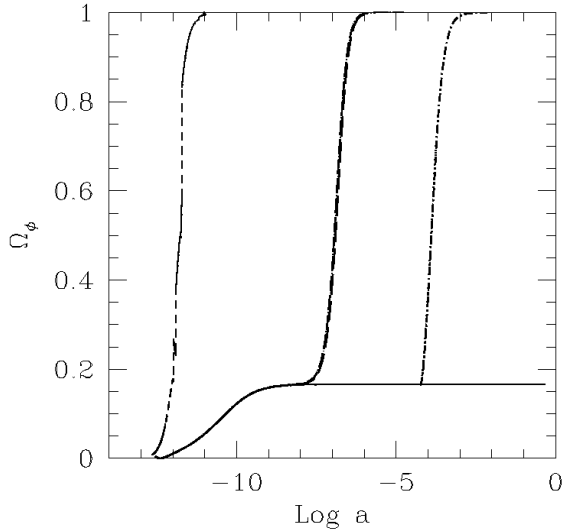}
\end{center}
\caption{$\Omega_{\phi}$ vs $\text{log a}$ for the generalized
function given by equation (\ref{oury}).  The solid line corresponds
to $c_{2}\neq 0$.  The dotted and dotted-short-dashed curves mixed
together correspond to $(c_{2},c_{1})$ and $(c_{2},c_{1},c_{0})$ as
the only non vanishing coefficients. The long-dashed curve corresponds
to $(c_{2}\neq0,c_{-1}\neq0)$ and the dotted-long-dashed one to
$(c_{2}\neq0,c_{-1}\neq0,c_{-2}\neq0)$. Finally the short-dashed curve
on the left corresponds to $c_{-2} \neq 0$.  }
\label{fig3}
\end{figure}

\begin{figure}[htbp]
\begin{center}
\includegraphics[scale=0.4]{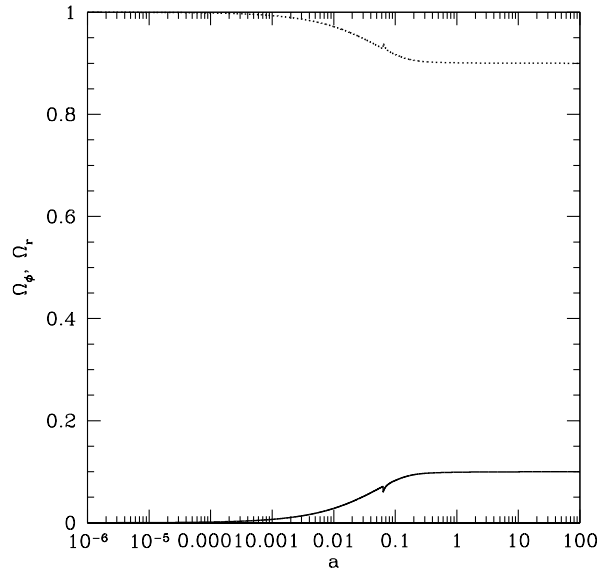}
\end{center}
\caption{A mixed case with $n=1$ and $n=3$ which exhibits a perfectly stable 
tracker, in the radiation era.
}
\label{mixed_one_three}
\end{figure}

In Fig.\ref{fig3} we plot the evolution of $\Omega_\phi$ in the
presence of nonrelativistic matter.  With only $c_{2}$ as nonzero
coefficient, we have an apparently stable attractor. If we choose in
(\ref{oury}) a nonzero $c_{1}$ (while keeping the nonzero $c_2$)
we find that the tracking is still
possible but less stable than the one with $c_{2}$ alone. Hence, we
can have longer lived tracking by decreasing $c_{1}$.  Adding $c_0$
with suitably chosen coefficients (see subsection on analytical
results below) does not change this behavior as can be seen in
Fig.\ref{fig3}. The curves corresponding to the two cases just
discussed essentially overlap, and depart from tracking  at $\log a=-4$.  
It seems that, in general, mixing cases of different $n$ together will
lead to the loss of tracking behavior, with at best having 
temporary trackers. While this turns out to be the most frequent 
behavior there are counter examples : a mixed case of $n=1$ and $n=3$ 
with $\mu_a = 5/7$ and $c_1/c_3 = - 4/49$ also exhibits perfectly 
stable tracker behavior (see fig.~\ref{mixed_one_three}). As we will see in the next section on
analytical treatment, the existence of tracking behavior or not,
can be understood in terms of a generic set of rules.

Tracking behavior disappears altogether when we add a negative power
in $({\hat \mu}-\mu_a)$.  
We can take either $c_{-1}\neq 0$ {\it or}
$c_{-1}\neq 0$ {\it and} $c_{-2}\neq 0$ and the effect will
essentially be the same.  The curves corresponding to these cases
almost overlap in Fig.\ref{fig3} (by chance), and $\Omega_\phi$ rapidly tends to $1$ at $\log a =-8$. In both
of these cases, the system exits tracking quite quickly.  The initial
stage tracking is a result of the presence of $c_{2}$; if we set
$c_{2}=0$ and keep $c_{-2} \neq 0$, for example, the leftmost curve of
Fig.\ref{fig3} indicates that tracking is impossible. 
 Higher powers of $({\hat \mu}-\mu_a)$ can suppress this instability
temporarily.

\begin{figure}[htbp]
\begin{center}
\includegraphics[scale=0.4]{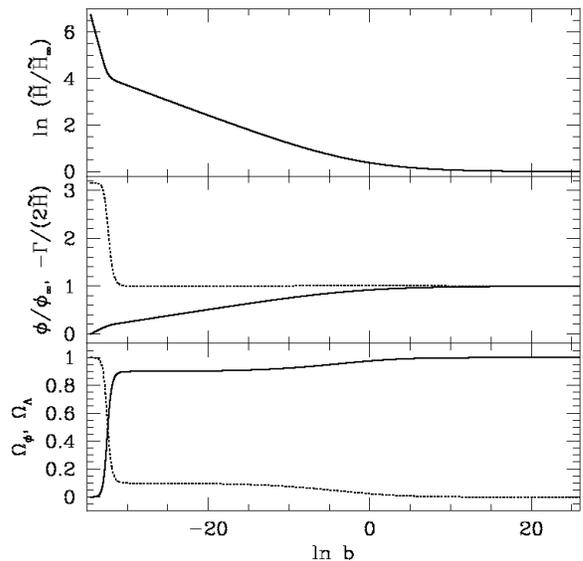}

\end{center}
\caption{A case with only $c_{-3}\ne 0$ and a cosmological constant. Upper panel : 
Evolution of $\ln(\tilde{H}/\tilde{H}_\infty)$ with $\ln b$ (see section on
the analytical treatment of the $n=-3$ case for the definition of $\tilde{H}_\infty$) . Middle panel :
Evolution of $\phi/\phi_\infty$ (solid) and $\Gamma/\tilde{H}$ (dotted) with $\ln b$  (see section on 
the analytical treatment of the $n=-3$ case for the definition of $\phi_\infty$). 
Lower panel : Evolution of the relative densities $\Omega_\phi$ (dotted) and $\Omega_\Lambda$ (solid) showing
an initial temporary tracking phase followed by $\Lambda$-domination.
}
\label{minus_three}
\end{figure}

What if we add more negative powers of $n$? The lower panel  of  Fig.\ref{minus_three} shows
the evolution of the relative densities $\Omega_i$ in the case where
only $c_{-3}\ne 0$ and where only a cosmological term is present. 

The middle panel of the same figure show that $\Gamma/\tilde{H}$ is a constant during tracking while
$\phi$ tends to a constant in the infinite future. The upper panel shows that $\tilde{H}$ also
tends to a constant.
In the case where only radiation and/or pressureless matter is present, we find that evolution
has an early unstable tracking behavior just like the $\Lambda$ case, but eventually 
evolution stops in finite time. Adding a cosmological
term can remedy the situation by creating a second tracking era.

\begin{figure}[htbp]
\begin{center}
\includegraphics[scale=0.4]{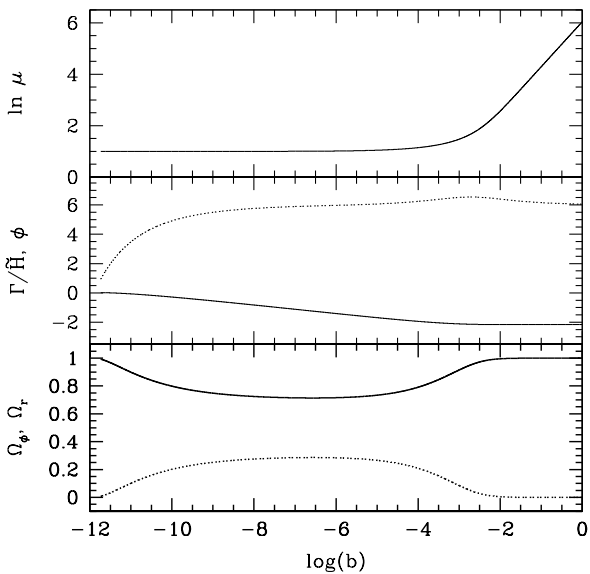}
\end{center}
\caption{A case with only $c_{-4}\ne 0$. 
Upper panel :
Evolution of $\ln\mu$ (solid) with $\ln b$. 
Middle panel : Evolution of $\phi$ (solid) and $\Gamma/\tilde{H}$ (dotted) with $\ln b$.
Lower panel :
 Evolution of relative densities of radiation (solid) and scalar field (dotted) versus $\ln b$ with only a radiation fluid
 present, showing a late radiation domination. 
}
\label{minus_four}
\end{figure}

Considering even more negative terms, e.g. only $c_{-4}\ne 0$ as in Fig.\ref{minus_four}
gives yet different behaviors. An early tracking era eventually
comes to an end followed by a fluid dominated era ($\Omega_X = 1$).
During both eras we find that the field $\Gamma$ always evolves
linearly with $\log t$ (if the background fluid is radiation).

Our numerical study of the dynamics of this system picks out four
types of behavior: perfect tracking, tracking followed by domination
of $\Omega_\phi$, no tracking and eventual singular behavior and 
finally initial tracking and eventual fluid domination.  Given that we have considered a reasonable
generalization of the function proposed by Bekenstein, we can see that
these types of behavior will be widespread in TeVeS. We can now
proceed to understand what lies behind these different regimes.

\section{Analytical Results}

In this section we investigate the existence and stability of
tracker solutions and develop approximate
analytical solutions to the evolution of the scalar field for the
various choices of function and tracking regimes.

When does the field track? The numerical studies of the previous section showed that
tracking occurs as $V'$ tends to its zero point. It therefore makes sense to
study solutions when $\mu$ is close to this point, i.e. we let 
$\hat{\mu} = \mu_a( 1 + \epsilon)$ with $\epsilon \ll 1$.
The impact of different terms in the function series is investigated bellow.

\subsection{Case $n\geq1$}
In this case the function $V$ and its derivative $V'$ are given by
\begin{eqnarray}
 V_{(n)} &=&  \frac{\mu_0}{16\pi\ell_B^2}  \bigg[ I_{n} +  \frac{n+2+\mu_a  +  (n+1)\hat{\mu} }{(n+1)(n+2) }(\hat{\mu} - \mu_a)^{n+1}
   \nonumber \\
&&  + \sum_{m=1}^n \frac{(1 - \mu_a)^{n-m}}{m} (\hat{\mu}-\mu_a)^m 
\nonumber \\
&&   + (1 - \mu_a)^n \ln|\hat{\mu} - 1|\bigg]
 \end{eqnarray}
 and
\begin{equation}
 V_{(n)}' =  \frac{\mu_0^2}{16\pi\ell_B^2}   \frac{\hat{\mu}^2}{\hat{\mu} -1}(\hat{\mu} - \mu_a)^n
 \end{equation}
respectively, where we put a subscript $(n)$ to denote the case considered, and where
$I_{n}$ is an integration constant (for the case $n$).

Clearly $V_{(n)}'(\mu_a)=0$ for any $n$. In the case of $V_{(n)}$ however, it depends on the
choice of the integration constant $I_{n}$. We will  discuss the relevance of this constant
further below.

Let us choose for the moment the constant $I_{n}$ such that  $V_{(n)}(\mu_a) = 0$.
We then expand the relevant quantities as a Taylor series about the point $\hat{\mu} = \mu_a$. We have
\be
\label{Vtaylor}
 V=\frac{\mu_0 \mu_{a} \hat{V}_{0}}{n+1} \epsilon^{n+1}
\ee 
and
\be 
\label{Vptaylor} 
 V'= \hat{V}_{0} \epsilon^n
\ee 
where $\hat{V}_0 =  \mu_0^2 \mu_a^{2+n} /(16\pi\ell_B^2 (\mu_a - 1) )$.
The field density is therefore given by
\be 
\rho_\phi =\frac{1}{16\pi G} e^{2\phi} \mu_0 \mu_a \hat{V}_0 \epsilon^n
\ee
to lowest order in $\epsilon$.

Using the constraint equation which links $\mu$ with $\dot{\phi}$, namely (\ref{constraint}), and 
taking the square root, we obtain 
\be
\label{phisko} 
\dot{\phi}=\beta \sqrt{\frac{\hat{V}_0}{2}}\;\; \epsilon^{n/2} 
\ee
where $\beta= \pm 1$ is an artifact of taking the square root.

The Taylor expansion of the auxiliary scalar field $\Gamma$  is 
\be
\Gamma=-2 \mu_{a} \mu_{0} \beta \sqrt{\frac{\hat{V}_0}{2}}\;\; \epsilon^{n/2} 
\ee

Since we are interested on finding possible tracking behavior, motivated by the
numerical results, it is reasonable to make the ansatz $\rho_\phi = A \rho_X$
with $A$ a constant, giving
 \be
\label{ansatz} 
\epsilon^{n}=A \frac{16 \pi G}{\mu_0 \mu_a\hat{V}_0 } e^{-2 \phi} \rho_X
\ee 
 Using the  Friedmann equation we get 
\be 
\label{friedsko} 
 \tilde{H}= \sqrt{\frac{8 \pi G(1 + A)}{3} e^{-2 \phi} \rho_X}, \ee 
Eliminating $\epsilon^n$ from the auxiliary scalar field $\Gamma$ we get
\be 
\label{gamskor}
\Gamma=-2 \beta \sqrt{ 8 \pi G A  \mu_{a} \mu_{0}  e^{-2 \phi} \rho_X}. 
\ee 
Using (\ref{evol_gamma})
 along with the derivative of   (\ref{gamskor}) we find
 \be
\Omega_\phi= \frac{A}{1+A} =\frac{(1+3w)^{2}}{3 \mu_a \mu_0(1-w)^2}. 
\ee
More specifically in the three special eras considered above,
namely radiation, matter and $\Lambda$ eras, we get
 $\Omega_{\phi}=\frac{1}{3 \mu_{a} \mu_{0}}$ in matter and $\Lambda$ eras, 
 while $\Omega_{\phi}=\frac{3}{\mu_{a} \mu_{0}}$ in the radiation era. 
 Taking $\mu_{a}=2$ for the Bekenstein toy model we obtain the results previously found in \cite{SMFB}.

Now we specify the constant $\beta$ which tells whether $\dot{\phi}$ is negative or positive.
Changing the time variable to $\ln a$ and using (\ref{friedsko}) and (\ref{phisko}), one finds that $\frac{d
\phi}{d \text{ ln} a}= \phi_{1}$ is approximately a  constant during tracking, given by
\be
\label{phio} 
\phi_{1}=\frac{1+3 w}{\mu_{a} \mu_{0}(1-w)\beta-(1+3 w)}
\ee
This results to $\phi=\phi_0 + \phi_{1} \text{ ln}a$. Introducing the above in (\ref{ansatz}) we find 
$\epsilon \propto a^{-\frac{2 \phi_{1}+3\left(1+w \right)}{n}}$
where $\rho_0$ is the fluid  density at $a=1$.
Consequently $\epsilon$ will
be a decreasing function going to $0$ asymptotically, and
$\dot{\epsilon} \rightarrow 0$ if and only if tracking is stable,
that's to say if $2 \phi_{1} +3\left(1+w\right)>0$. Using
 (\ref{phio}) this condition is equivalent to 
 \be \frac{-\left(1+3 w\right)^{2}+3\left(1-w^{2}\right) \mu_{a} \mu_{0}
\beta}{\mu_{a} \mu_{0}\left(1-w\right) \beta-\left(1+3w\right)}>0 \ee 
The above inequality is satisfied if and only if 
either $\beta = 1$ and $w>-1/3$ or $\beta = -1$ and $w<-1/3$.
Therefore $\beta$ is completely fixed by the equation of state of the background fluid. 
In particular we see that $\dot{\phi}$ has to change sign when passing from
the matter to $\Lambda$ era which results to $\rho_\phi$ going momentarily
to zero~\cite{SMFB}.

Now let us discuss the role of the integration constant $I_{n}$.
Its sole impact would be to add a constant in the Friedmann equation.
What can motivate such a constant? For exact MOND limit we need $V\rightarrow 0$ as $\mu \rightarrow 0$.
It is in general impossible to have both $V(0)  = 0$ and $V(\mu_a) = 0$, except in a very special
and unique case : the Bekenstein toy model, i.e. $n=2$ and $\mu_a = 2$. In every other case (not a mixed case
with many $c_n$ nonzero, which will be consider in another section), setting $V(\mu_a)=0$ would spoil
the \textit{exact} MOND limit. The impact of such a constant in the evolution equations however, is what 
precisely destroys the perfect tracker. In other words as long as $I_{n} \ll 16\pi G e^{-2\phi} \rho_X$,
we would have the tracker solution found above. When eventually $\rho_X$ decreases below this threshold,
which is bound to happen always, tracking stops. In such case there are two possible behaviors after 
tracking stops. If $V(\mu_a) < 0 $ the requirement that the LHS of the Friedmann equation is positive,
creates a pathological situation where $\Omega_\phi$ decreases without bound \textit{below} zero
while the fluid relative density increases unbounded above one to counterbalance.  In the opposite case where $V(\mu_a) > 0 $ the universe enters a 
 de-Sitter phase in the Einstein frame where $\dot{b}/b$ tends to a constant, and $\Omega_\phi$ tends 
 rapidly to $1$. This second case is what stops the temporary trackers found in the previous section
 numerically, in the mixed cases.

We end this case by noting that the Hubble parameter in our approximation will
be slightly different from the one we obtain in General Relativity but rather if
the background radiation temperature is $T$, it will be given by $H \propto T^\gamma$ where $\gamma = 2\phi_1 + \frac{3}{2}(1+w)$. 
This can be potentially important when calculating nucleosynthesis constraints, as
in the radiation era we would have $\gamma = 2(1 + \phi_1)$ rather than the
canonical value $\gamma=2$. The impact of this kind cosmologies on  nucleosynthesis has been
investigated in ~\cite{CK}.

\subsubsection{Case $n=0$}
In this case the function is given by 
\be 
V_{(0)} =\frac{\mu_0^3}{16\pi\ell_B^2}\left( I_0 + \frac{1}{2}\hat{\mu}^2 + \hat{\mu} + \ln |\hat{\mu} - 1|\right)
\ee
and
\be 
\label{potv2} 
V'_{(0)} = \frac{\mu_0^2}{16\pi\ell_B^2} \frac{\hat{\mu}^{2}}{\hat{\mu}-1}.
\ee
The constraint can be inverted analytically to give
\begin{equation}
\hat{\mu} = \frac{4\sqrt{\pi}\ell_B}{\mu_0^2}\dot{\phi} \left[ 4\sqrt{\pi}\; \ell_B \dot{\phi} + \sqrt{16\pi\ell_B^2\dot{\phi}^2 - 2\mu_0^2} \right]
\end{equation}
It is therefore clear that $|\dot{\phi}|$ is bounded from below : $|\dot{\phi}|\ge \mu_0 / (\sqrt{8\pi}\ell_B)$. At this value $\hat{\mu}$ takes its
minimum possible value given by $\mu_c = 2$. The upshot is that contrary to the  
case where $n\ge1$, we have $V'(\mu_c) = \mu_0^2/(4\pi \ell_B^2)$ i.e. it is not zero. We will show in this subsection
that this creates a pathological situation.

First note that since $\dot{\phi}$ cannot be zero, the two branches of positive and negative
$\dot{\phi}$ are disconnected and we can change variables from $\dot{\phi}$ to $\hat{\mu}$. 
The equivalent of (\ref{evol_gamma}) is then given by
\begin{equation}
\label{mudot}
 \frac{3\hat{\mu} - 4}{\hat{\mu} - 1} \dot{\hat{\mu}}  + 6 \tilde{H}\hat{\mu} + \frac{4\beta(1+3w) \sqrt{2\pi}\ell_B}{\mu_0^2} \frac{\sqrt{\hat{\mu} - 1}}{\hat{\mu}} y
  = 0
\end{equation}
where $y = 8\pi G e^{-2\phi} \rho_X$ and $\beta$ is once again the sign of $\dot{\phi}$.

Suppose that the $\hat{\mu}=2$ point is reached at
some moment in time $t_c$. Then $V(\hat{\mu}=2) =  \mu_0^3 ( I_0 + 12)/(16\pi\ell_B^2)$ giving
 \begin{equation}
 \dot{\hat{\mu}}_c  =  - \sqrt{ \frac{3\mu_0^3(I_0+12)}{8\pi\ell_B^2} + 12 y_c}  - \frac{2\beta(1+3w)\sqrt{\pi}\ell_B}{\mu_0^2\sqrt{2}} y_c 
 \end{equation}
 where the subscript "$c$" on any variable denotes the value of that variable at time $t_c$, and where we have 
 assumed that $\dot{b}/b > 0$. 
We consider again two subcases depending on the sign of $\beta (1 + 3w)$. 

If $\beta (1 + 3w)>0$, then $\dot{\hat{\mu}}$ is 
always negative (including the point $\dot{\hat{\mu}}_c$) except the very special case where $I_0 = -12$  and $y_c = 0$.
Since however we are free to choose the initial condition for $\mu$ and $y$ independently, we will get that
generically $\dot{\hat{\mu}}_c<0$. Therefore the system of differential equations is bound to become ill-defined at some
point $t_c$. 

If  $\beta (1 + 3w)<0$ then there are values of $\dot{\hat{\mu}}$ (and also $\dot{\hat{\mu}}_c$) which are positive given a large enough
$y$, which means that $\hat{\mu}$ will bounce off the point $\hat{\mu}=2$ and start increasing again. However as the equations are 
propagated further in time, $y$  decreases even further until once again  $\dot{\hat{\mu}}<0$ and $\hat{\mu}$ will decrease back to
$\hat{\mu} = 2$. This time though  $\dot{\hat{\mu}}_c<0$, and the system of differential equations becomes once more ill-defined.

\subsubsection{Case $n=-1$}
It turns out that for negative $n$ we need to consider each case separately for the first few $n$'s. Lets start with the
$n=-1$ case.

In this case the function is given by 
\begin{eqnarray}
V_{(-1)} &=& \frac{\mu_0^3}{16\pi\ell_B^2} \bigg( I_{-1} + \hat{\mu} + \frac{\mu_a^2}{\mu_a - 1} \ln|\hat{\mu} - \mu_a| \nonumber \\
  && -  \frac{1}{\mu_a  - 1}  \ln |\hat{\mu} - 1|\bigg)
\end{eqnarray}
and
\be 
V'_{(-1)} = \frac{\mu_0^2}{16\pi\ell_B^2} \frac{\hat{\mu}^{2}}{(\hat{\mu}-1)(\hat{\mu} - \mu_a)}.
\ee

The constraint can be inverted analytically to give
\begin{equation}
\hat{\mu} = \frac{(\mu_a + 1)\dot{\phi}^2 + |\dot{\phi}| \sqrt{(\mu_a-1)^2\dot{\phi}^2 + \frac{\mu_0^2 \mu_a}{8\pi\ell_B^2}}}{2\dot{\phi}^2 - \frac{\mu_0^2}{16\pi\ell_B^2}}
\end{equation}
Restricting the allowed range for $\hat{\mu}$ to be greater than $\mu_a$ (since we have a singularity in $V'$ at $\hat{\mu} = \mu_a$) we see that $\dot{\phi}$ is once again
bounded from below as $|\dot{\phi}|\ge \mu_0/(4\ell_B \sqrt{2\pi})$, similarly to the $n=0$ case.
Since the two branches of positive and negative $\dot{\phi}$ are once again disconnected, 
we can change variables from $\dot{\phi}$ to $\hat{\mu}$ as in the case $n=0$.

The equivalent of (\ref{evol_gamma}) is then given by
\begin{eqnarray}
\dot{\hat{\mu}} &=& -\frac{2(\hat{\mu}-1)(\hat{\mu}-\mu_a)}{2\hat{\mu}^2 - 3(\mu_a + 1)\hat{\mu} + 4\mu_a} \bigg[
3 H \hat{\mu}
\nonumber \\
  && 
   + \frac{4\beta(1+3w)\sqrt{\pi}\ell_B}{\mu_0^2\sqrt{2}} \frac{\sqrt{(\hat{\mu}-1)(\hat{\mu}-\mu_a)}}{\hat{\mu}} y\bigg]
\label{mudotneg}
\end{eqnarray}
where $y = 8\pi G e^{-2\phi} \rho_X$ and $\beta$ is once again the sign of $\dot{\phi}$.
Now for $\hat{\mu}\ge \mu_a$ we have that $2\hat{\mu}^2 - 3(\mu_a + 1)\hat{\mu} + 4\mu_a = 0$ at
a value $\hat{\mu} = \mu_c = \frac{1}{4}\left[3(\mu_a + 1) + \sqrt{9\mu_a^2 - 14\mu_a + 9} \right]$.
In other words there the evolution of $\dot{\hat{\mu}}$ reaches a singularity at $\hat{\mu} = \mu_c$. 
To make things worse, since $\dot{\hat{\mu}}$ is negative for $\hat{\mu}>\mu_c$ while being positive
for $\hat{\mu}<\mu_c$ then this singularity is always reached! This is true independently of the sign of $\beta$ for the
same reason as the case $n=0$, i.e. for positive $\tilde{H}$, as $y$ decreases below some threshold value the term 
proportional to $y$ can be neglected.

\subsubsection{Case $n=-2$}
Let us now discuss the $n=-2$ case.

In this case the function is given by 
\begin{eqnarray}
V_{(-2)} &=& \frac{\mu_0^3}{16\pi\ell_B}\bigg( I_{-2} + \frac{\mu_a^2}{1-\mu_a} \frac{1}{\hat{\mu} - \mu_a}
  +  \frac{\mu_a(\mu_a - 2)}{(\mu_a-1)^2} \ln|\hat{\mu} - \mu_a| \nonumber \\
  && +  \frac{1}{(\mu_a  - 1)^2}  \ln |\hat{\mu} - 1|\bigg)
\end{eqnarray}
and
\be 
V'_{(-2)} = \frac{\mu_0^2}{16\pi\ell_B} \frac{\hat{\mu}^{2}}{(\hat{\mu}-1)(\hat{\mu} - \mu_a)^2}.
\ee

Contrary to the last case ( $n=-1$), $\dot{\phi}$ is no longer bounded from below and can
take all possible values. However the point $\dot{\phi}=0$ can only occur at $\mu\rightarrow \infty$,
which means that once again the two sectors of positive and negative $\dot{\phi}$  are disconnected, since
the $\mu\rightarrow \infty$ is a singular point (the Friedman equation blows up).
We can therefore follow the approach we took in the $n=-1$ case and change variables from $\dot{\phi}$ to $\mu$ giving
\begin{eqnarray}
\dot{\hat{\mu}} &=& - \frac{2(\hat{\mu} - 1)(\hat{\mu} - \mu_a)}{\hat{\mu}^2 - (3\mu_a + 2)\hat{\mu} + 4\mu_a } \bigg[
3 \mu_0 \hat{\mu}   H 
\nonumber \\
  && 
+ \frac{4\beta(1+3w)\sqrt{\pi}\ell_B (\hat{\mu} - \mu_a) \sqrt{\hat{\mu}-1}} {\mu_0 \hat{\mu}\sqrt{2}}y\bigg]
\label{mudotneg2}
\end{eqnarray}
where $y = 8\pi G e^{-2\phi} \rho_X$ and $\beta$ is once again the sign of $\dot{\phi}$.
We immediately see that the same problem as the $n=-1$ case arises : there is a singularity in the
evolution when $\hat{\mu} = \mu_c = \frac{1}{2}\left[ 3\mu_a + 2 + \sqrt{9\mu_a^2 - 4\mu_a + 4}\right]$. Since $\mu_c > \mu_a$
always while at the same time
 $\dot{\mu}<0$ for $\hat{\mu}>\mu_c$ and  $\dot{\mu}>0$ for $\hat{\mu}<\mu_c$ , then once again this singularity is always reached.

\subsubsection{Case $n=-3$}
We now turn to the $n=-3$ case.

In this case the function is given by 
\begin{eqnarray}
V_{(-3)} &=& \frac{\mu_0^3}{16\pi\ell_B} \bigg( I_{-3} + \frac{\mu_a^2}{2(1-\mu_a)} \frac{1}{(\hat{\mu} - \mu_a)^2} \nonumber \\
  && - \frac{\mu_a(\mu_a-2)}{(1-\mu_a)^2}  \; \frac{1}{\hat{\mu} - \mu_a}  \nonumber \\
  && +  \frac{1}{(1- \mu_a)^3}  \ln |\frac{\hat{\mu} - 1}{\hat{\mu} - \mu_a}|\bigg)
\end{eqnarray}
and
\be 
V'_{(-3)} = \frac{\mu_0^2}{16\pi\ell_B}  \frac{\hat{\mu}^{2}}{(\hat{\mu}-1)(\hat{\mu} - \mu_a)^3}.
\ee

Contrary to the $n=-1$ and $n=-2$ cases, in this case, $2V' + \mu V''$ never goes to zero for $\hat{\mu}>\mu_a$,
hence no singularity of the same type as $n=-1$ and $n=-2$ occurs in this case.
We consider two limiting cases of behavior. 

The first case is when $\hat{\mu}$ approaches $\mu_a$. In this limit
we get that $V'$ is of order $\dot{\phi}^2$ while $V$ is of lower order, 
giving  $\mu V' + V = 2 \mu_0 \mu_a \dot{\phi}^2$. Therefore in this limit the equations take the form
\begin{equation}
	3\tilde{H}^2 = \mu_0 \mu_a \dot{\phi}^2 + 8\pi G e^{-2\phi} \rho_X 
\end{equation}
and
\begin{equation}
	\ddot{\phi} + 3\tilde{H}\dot{\phi} + \frac{(1+3w) }{2\mu_0 \mu_a} 8\pi G e^{-2\phi} \rho_X =0
\end{equation}
in other words the system behaves as a one with a canonical scalar field coupled to matter. In this limit
we have tracker solutions such that $\dot{\phi} = -\frac{1+3w}{(1-w)\mu_0 \mu_a} \tilde{H}$,
 which gives $\Omega_\phi = \frac{(1+3w)^2}{3(1-w)^2\mu_0 \mu_a}$.
This tracker however is unstable (but can be long-lived depending on
the fluid energy budget), i.e. as $\mu$ becomes larger this limit ceases to be valid.

The second case is when $\mu\rightarrow \infty$. We expand all functions in powers of $\epsilon = 1/\hat{\mu}$, giving
$V' = \mu_0^2\epsilon^2 \left[ 1 + (3\mu_a + 1)\epsilon\right]/(16\pi\ell_B^2)$ resulting to lowest order in $\epsilon$, 
$\dot{\phi} = \beta \mu_0 \epsilon/(4\ell_B\sqrt{2\pi})$, and $\Gamma = -\beta\mu_0^2 \sqrt{2}\left[ 
1 + (3\mu_a + 1)\epsilon/2\right]/(4\ell_B\sqrt{\pi})$. It also turns out that although $\mu V'$ is of $O(\epsilon)$, this order is
precisely canceled from a term in $V$, resulting to $\mu V' + V = \mu_0^3  (3\mu_a + 1) \epsilon^2/(32\pi\ell_B^2)$ 
to lowest order in $\epsilon$.
The evolution equations become
\begin{equation}
 3\tilde{H}^2 = \frac{\mu_0^3}{64\pi\ell_B^2}  (3\mu_a + 1) \epsilon^2 + y
\end{equation}
and
\begin{equation}
  \frac{3\mu_a+1}{2}\dot{\epsilon} + 3 \left( 1 + \frac{3\mu_a+1}{2}\epsilon\right) \tilde{H}
 + \frac{4(1+3w)\beta \ell_B \sqrt{\pi}y}{\mu_0^2 \sqrt{2}} =0 
\label{eps_dot_m3}
\end{equation}
First we consider the case when the fluid is a cosmological constant. Letting $\epsilon$ and $\dot{\epsilon}$ to go
zero, we get that the Hubble parameter tends to a constant $\tilde{H} \rightarrow \tilde{H}_\infty= \mu_0^2\sqrt{2}/(8\sqrt{\pi}\ell_B)$, while
$\phi\rightarrow \phi_\infty = \frac{1}{2}\ln{\frac{8\pi G \rho_\Lambda}{3\tilde{H}_\infty^2}}$ and $\beta=1$. Perturbing
$\phi$ about $\phi_\infty$ as $\phi = \phi_\infty( 1 + \delta )$ we get that $\tilde{H}^2 = \tilde{H}_\infty^2(
1 - 2 \phi_\infty \delta)$. Differentiating  (\ref{eps_dot_m3}) once under these approximations and eliminating
$\delta$ via $2\mu_0\phi_\infty \dot{\delta} = \tilde{H}_\infty\epsilon$ we get
\begin{equation}
  \ddot{\epsilon} + 3  \tilde{H}_\infty \dot{\epsilon} + \frac{6 \tilde{H}_\infty^2 }{(3\mu_a+1)\mu_0} \epsilon =0 
\end{equation}
This has decaying solutions of the form $\epsilon = \epsilon_1 e^{-A_{+}t} + \epsilon_2  e^{-A_{-}t}$ where
\begin{equation}
A_{\pm} = \frac{3}{2}\tilde{H}_\infty\left( 1 \pm \sqrt{1 - \frac{8}{3(3\mu_a+1)\mu_0 }}\right)
\end{equation}
provided that $(3\mu_a+1)\mu_0\ge8/3$. The final stage of this particular case is a universe with $\Omega_\Lambda = 1$.

It turns out that when the fluid is not a cosmological constant then, the evolution reaches
$\epsilon \rightarrow 0$ in finite time. This is once again a singularity of a similar form to the 
$n=0$ case. The reason we have such a singularity is the same as the $n=0$ case : one can choose 
initial conditions such that $\dot{\phi} = 0$ which corresponds to $\mu\rightarrow \infty$, at any point
in time. However this is inconsistent with the equations of motion except in the special case of the 
presence of a cosmological constant. Therefore in this case, if the evolution is to be nonsingular
in the infinite future, one must include a cosmological constant.

\subsubsection{Case $n\le-4$}
Finally we consider the $n\le-4$ case.

In this case the function is given by 
\begin{eqnarray}
V_{(-n)} &=& \frac{\mu_0^3}{16\pi\ell_B^2} \bigg[ I_{-n}  + \frac{\mu_a^2}{(n-1)(1 - \mu_a)} \frac{1}{(\hat{\mu} - \mu_a)^{n-1}}  \nonumber \\
  && - \frac{\mu_a(\mu_a - 2)}{(n-2)(1 - \mu_a)^2} \frac{1}{(\hat{\mu}-\mu_a)^{n-2}} \nonumber \\
&& +  \frac{1}{(1- \mu_a)^n} \sum_{m=1}^{n-3} \frac{(1-\mu_a)^m}{m (\hat{\mu}-\mu_a)^m}  \nonumber \\
&& +  \frac{1}{(1- \mu_a)^n}  \ln |\frac{\hat{\mu} - 1}{\hat{\mu} - \mu_a}|\bigg]
\end{eqnarray}
and
\be 
V'_{(-n)} = \frac{\mu_0^2}{16\pi\ell_B^2} \frac{\hat{\mu}^{2}}{(\hat{\mu}-1)(\hat{\mu} - \mu_a)^n}.
\ee

It is easy to show that $2V' + \mu V''\ne0$ for $\hat{\mu}>\mu_a$ unless $\hat{\mu}\rightarrow \infty$. We therefore expect to
have smooth evolution and furthermore we expect that $\hat{\mu}$ will eventually increase to infinity. 

Eventually $\mu$ increases  to large values where we can 
 expand $V$ and $V'$ in powers of $\epsilon = 1/\hat{\mu}$. For $V'$ we get
\be
  V' \approx \frac{\mu_0^2}{16\pi\ell_B^2}  \epsilon^{n-1}
\ee
giving
\be
  \Gamma \approx \beta \mu_0^2 \frac{\sqrt{2}}{4\sqrt{\pi}\ell_B} \epsilon^{\frac{n-3}{2}}
 \label{Gammaepsilon1}
\ee
where $\beta$ is the sign of $\Gamma$.

 It turns out that
$\mu V' + V$ is at least of $O(\epsilon^2)$ for $n=-4$ or higher for $n<-4$ while
$\dot{\phi}$  is at least of $O(\epsilon^{3/2})$ for $n=-4$ or higher for $n<-4$.
 We can therefore assume that
the universe evolves like Einstein gravity with $\tilde{H}= H$ and we have the usual Friedmann equation : $3 H^2 = 8\pi G \rho_X$

We consider three separate subcases : $w=-1$, $w=1$ and $-1 < w < 1$.

For the $w=-1$ case we have that $H$ will tend to a constant.
The evolution equation for $\Gamma$ reads 
\be
\dot{\Gamma} + 3 H \Gamma + 6 H^2 = 0
\ee
The solution is
\be
 \Gamma = 2 H ( e^{-3Ht} - 1)
\ee
and therefore $\Gamma \rightarrow 0 $ as required by consistency.
Feeding back to (\ref{Gammaepsilon1}) we get that $\beta = -1$.

For the $w=1$ case we have that $H = \frac{1}{t}$ which means that $\Gamma$ evolves according to
\be
\dot{\Gamma} + \frac{3}{t} \Gamma - \frac{12}{t^2}  = 0
\ee
The solution in this case is
\be
 \Gamma = \frac{6}{t} + \frac{\Gamma_0}{t^3}
\ee
where $\Gamma_0$ is a constant. Once again $\Gamma \rightarrow 0 $ as required by consistency.
Feeding back to (\ref{Gammaepsilon1}) we get that $\beta = 1$.

The last case is when $-1 < w < 1$. In this case $\Gamma$ evolves as
\be
\dot{\Gamma} + 3 H \Gamma - 3(1+3w) H^2 = 0
\ee
which has solution
\be
\Gamma = \frac{2(1+3w)}{1-w} H = \frac{4(1+3w)}{3 (1-w^2)} \; \frac{1}{t}
\ee
and therefore $\Gamma$  consistently goes to zero.
Feeding back to (\ref{Gammaepsilon1}) we get that $\beta = 1$ for $w>-1/3$ and $\beta=-1$ for $w<-1/3$.

\subsection{Mixed cases}

Mixing cases for $n\ge1$ can have two generic kinds of behavior 
depending on whether $V(\mu_a) = 0$ or not. 

If $V(\mu_a) = 0$ then we have perfect trackers.
If $V(\mu_a) > 0$ there will be a temporary tracker until 
   $V(\mu_a) \sim 16\pi G e^{-2\phi} \rho_X$ after which the tracker is 
   destroyed and we get $\phi$ domination. The solution in the 
   $\phi$-domination era is de-Sitter space (in the Einstein frame).

Now the condition of exact MOND means that only a very restricted 
set of function can have $V(\mu_a) = 0$. For the single $n$  cases
only the Bekenstein toy model ($n=2$ and $\mu_a=2$) has this 
feature. By mixing cases however, this feature can arise again
for very specific choices of $\mu_a$ and $n$'s. For example 
this is impossible for a mixed $n=1$ and $n=2$ case, but possible
for a mixed $n=1$ and $n=3$ case with $\mu_a=2/7$ and $c_1/c_3 = -4/49$.

While individually $n=-1$ and $n-2$ lead to singularities,
mixing the two can for some choices of $c_{-1}$ and $c_{-2}$ 
lead to $2 V' + \mu V''$ being nonzero for all $\hat{\mu}>\mu_a$.
This will lead to a better behaved evolution although
problems can still persist as $\mu\rightarrow \infty$ just like
the $n=-3$ case. This last problem could be removed by mixing in
a positive power of $n$.

Mixing $n=0$ with positive $n$ alone cannot remove the problem when $\hat{\mu}\rightarrow\mu_a$,
i.e the evolution will still become ill defined at that point. One can still
however mix $n=0$ with both positive and negative powers; the effect of the 
negative power is to drive $\hat{\mu}$ away from the $\hat{\mu}=\mu_a$ point.

Finally the most general mixed case including both positive and negative
powers with suitably chosen coefficients $c_i$, will give a function $V'$ which goes to zero not at $\hat{\mu} = \mu_a$ 
but at a shifted point $\hat{\mu} = \mu_a'$. the large $\mu$ behavior 
will be dominated by the positive powers of $n$ while the small $\mu$
 behavior by the negative powers with the function still being singular
at  $\hat{\mu} = \mu_a$. If we restrict the evolution for values
$\hat{\mu}>\mu_a'$, then we have a situation similar to the positive $n$ cases. We therefore
expect to have the same behavior as if we had a function expanded about $\hat{\mu}=\mu_a'$
with positive powers only, i.e. we expect to get stable trackers or in the case of a nonzero
integration constant in $V$ eventual $\phi$-domination.

\section{Conclusions}
In this paper we have tried to extract general properties of the
cosmology of TeVeS gravity. We have restricted ourselves to analyzing
the background equations for a homogeneous and isotropic universe. We
have considered a general function for $\mu$ which we believe
encompasses most of the possible regimes one could encounter in this
theory.

We find that Bekenstein's choice of $V(\mu)$ naturally leads to stable tracking 
behavior. This confirms the
results of \cite{SMFB}. Generalizing
the Bekenstein function by including arbitrary strictly positive powers of $(\hat{\mu}-\mu_a)$ we
either get exactly the same type of tracking or temporary tracking followed
by $\phi$-domination. What controls this kind of behavior is the value of the
integration constant in $V(\mu)$. If $V(\mu_0 \mu_a) = 0$ then the tracking behavior
is retained to the infinite future, otherwise the tracking period is temporary followed
by $\phi$-domination. 

We find that the $n=0$, $n=-1$ and $n=-2$ cases alone lead to future singularities in the
evolution. The same happens in the $n=-3$ case unless the background fluid is a cosmological
constant. For $n\le4$ the evolution is nonsingular and eventually leads to fluid domination
with $\mu\rightarrow\infty$.
\begin{figure}[htbp]
\begin{center}
\includegraphics[scale=0.35]{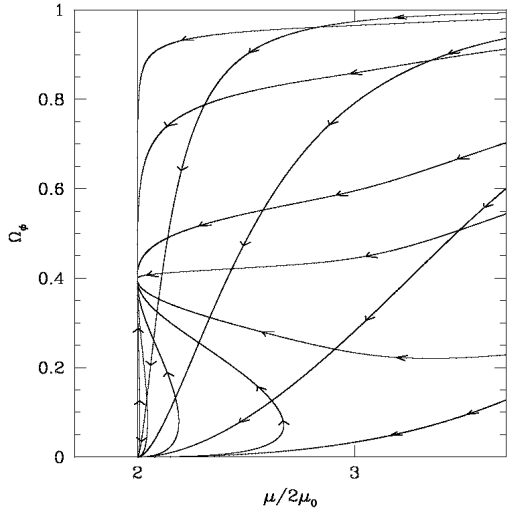}
\end{center}
\caption{Phase plot for the Bekenstein's model. All the curves join
the tracker solution ($\Omega_{\phi}$=0.4) and tracking remains
stable.}
\label{fig5}
\end{figure}


\begin{figure}[htbp]
\begin{center}
\includegraphics[scale=0.35]{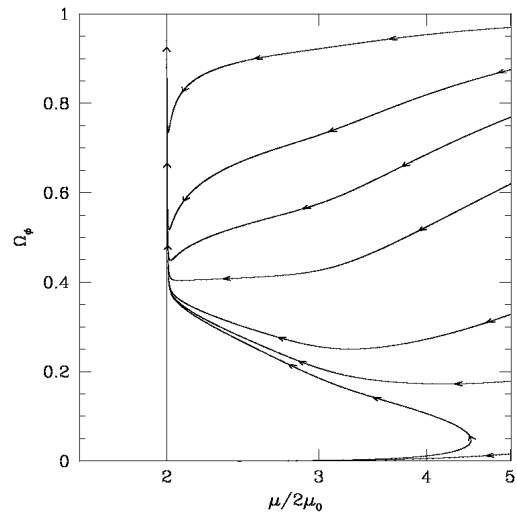}
\end{center}
\caption{Phase plot for the $n=2$ and $n=-1$ model. The upper curves
don't join the tracker solution ($\Omega_{\phi}$=0.4) and tracking
becomes unstable eventually leading to the singularity mentioned
in the text (in this case with a  $\phi$ dominated Universe). We used
$c_{2}c_{-1}<0$.}
\label{fig7}
\end{figure}

Finally, the singularities can be avoided by mixing cases together. If
positive and negative cases are mixed, it turns out that this is 
equivalent to a positive-only mixture but with a different expansion
parameter $\mu_a'$ rather than $\mu_a$, i.e. the 
function is equivalent to expanding in terms of positive
powers of  $(\hat{\mu} - \mu_a')$ in which case we again get the tracking
behavior discussed above.

A useful way to visualise the results of this paper is through the
 two phase plots in Figs.\ref{fig5},
and \ref{fig7}.  For
 each plot we pick a range initial conditions and we look at the
 function $\Omega_{\phi}(\hat{\mu})$ for the choice of $\mu_0$ such
 that $\Omega_{\phi}=0.4$ in the tracker limit.  In Fig.5
 (Bekenstein's model) we see that all the curves join the tracker
 solution.
 In Fig.\ref{fig7} we plot a model with $n=2$ and
 $n=-1$. With some choices of initial conditions we have a temporary
 tracker due to the $n=2$ factor and eventually rolling to the
 singularity mentioned in the section on the $n=-1$ case. While with
 other choices of initial conditions, the tracker is avoided all
 together.

The next step in the analysis of the cosmology of TeVeS is to
proceed with the programme initiated in \cite{SMFB}. In that work, it was shown that,
for Bekenstein's choice of function that it was possible to fit
observations of large scale structure and fluctuations in the cosmic
microwave background (CMB), the later if massive neutrinos of mass $\sim 2eV$ are present. 
This neutrino mass is on the border of  current neutrino mass experiments~\cite{numass} but not yet ruled out.
It has been claimed that these theories predict low odd peaks in the angular power spectrum of the CMB due
to the absence of dark matter \cite{MSS,WMAP3}. This is incorrect:
as explained in \cite{SMFB} (and subsequently reconfirmed in \cite{DL})
the extra fields in TeVeS sustain the growth of perturbations through recombination, driving the acoustic
oscillations in a manner entirely analogous to that of dark matter. Whether the power spectra based on 
 a generic function along the lines considered in this work can fit current observations is left for a future work.

\acknowledgments FB thanks C. Bourliot and C. Bourliot for support.
DFM acknowledges  the Humboldt Foundation, the Research Council of Norway through project number
159637/V30 and the Perimeter Institute for Theoretical Physics.
Research at Perimeter Institute is supported in part by the Government of Canada through NSERC and by the Province of Ontario through MEDT.

\end{document}